\documentclass[12pt]{article}%
\usepackage{amsfonts}
\usepackage{amsmath}
\usepackage{amssymb}
\usepackage{graphicx}%
\providecommand{\U}[1]{\protect\rule{.1in}{.1in}}
\newtheorem{definition}{Definition}

\begin{document}

\title{Data-Driven Fuzzy Modeling Using Deep Learning}
\author{Erick de la Rosa , Wen Yu\\{\small Departamento de Control Automatico}\\{\small CINVESTAV-IPN (National Polytechnic Institute)}\\{\small Mexico City, 07360, Mexico}}
\date{}
\maketitle

\begin{abstract}
Fuzzy modeling has many advantages over the non-fuzzy methods, such as
robustness against uncertainties and less sensitivity to the varying dynamics
of nonlinear systems. Data-driven fuzzy modeling needs to extract fuzzy rules
from the input/output data, and train the fuzzy parameters.

This paper takes advantages from deep learning, probability theory, fuzzy
modeling, and extreme learning machines. We use the restricted Boltzmann
machine (RBM) and probability theory to overcome some common problems in data
based modeling methods. The RBM\ is modified such that it can be trained with
continuous values. A probability based clustering method is proposed to
partition the hidden features from the RBM, and extract fuzzy rules with
probability measurement. An extreme learning machine and an optimization
method are applied to train the consequent part of the fuzzy rules and the
probability parameters. The proposed method is validated with two benchmark problems.

\end{abstract}

\section{Introduction}

Fuzzy modeling uses a set of IF-THEN rules to represent a nonlinear system.
Each rule only model the local dynamic of the system. A fuzzy model can
approximate a large class of nonlinear systems, while keeping linguistic
propositions of human thinking \cite{Zadeh}. Moreover, the fuzzy model can be
regarded as an universal estimator. It can approximate any nonlinear function
to any prescribed accuracy, provided that sufficient fuzzy rules are available
\cite{Brown}\cite{Lin}. It is often claimed that fuzzy models are more robust
than nonfuzzy methods against the sensitivity of variations of the data, or
varying dynamics of nonlinear systems \cite{Kumar}.

Data-driven fuzzy modeling uses observed data to construct a fuzzy model
automatically. It needs two processes: 1) extracting suitable fuzzy rules from
the data and deriving a fuzzy model; 2) updating the parameters of the fuzzy
model with the data. The first process is called structure identification, the
second process is called parameter identification. The key problem of the
structure identification is the extraction of the fuzzy rules. The fuzzy rules
can be obtained from mechanistic prior knowledge of nonlinear systems
\cite{Leski}, from the knowledge of experts \cite{Brown}, or from data
\cite{Lin}\cite{Zhang}. However, it is difficult to obtain mechanistic prior
knowledge for many nonlinear processes, and the expert method needs the
un-bias criterion and the trial-and-error technique \cite{Rivals}, which can
only be applied off-line. The data-driven fuzzy modeling is very effective to
identify a wide class of complex nonlinear systems when we have no complete
model information, or even when we consider the nonlinear system as a black
box \cite{Noori}.

Extraction of fuzzy rules from the input/output data usually uses the
partition method, which is also called fuzzy grid \cite{Lam}. Many data
clustering methods are applied for structure identification, such as fuzzy
C-means clustering \cite{Mitra}, mountain clustering \cite{Mitra}, and
subtractive clustering \cite{Chiu}. These approaches require that the data is
ready before the modeling. On-line clustering with a recursively calculated
spatial proximity measure is given in \cite{Angelov}. The combination of
on-line clustering and genetic algorithms for fuzzy systems is proposed in
\cite{Juang2}. In \cite{Yu1} the input space is automatically partitioned into
fuzzy subsets by adaptive resonance theory. Besides these clustering
approaches, fuzzy rule extraction can also be realized by neural networks
\cite{Juangb}, genetic algorithms \cite{Rivals}, singular-value decomposition
\cite{Chiang} and support vector machines \cite{Cristianini}. These data based
clustering methods do not use the probability distribution information of the data.

In the sense of probability theory, the objective of system modeling is to
obtain a conditional probability distribution $P(y|\mathbf{x})$ \cite{Erhan1},
where $\mathbf{x}$ is the input and $y$ is the output. Recent results show
that deep learning techniques can learn the probability distribution $P(x)$ of
the input space with an unsupervised learning method. \cite{Bengio1} shows
that in the unsupervised learning stage, the input information are sent to
hidden layers to construct useful statistical features. This mechanism
improves the corresponding input/output representation. The input distribution
$P(x)$ appears in the hidden units via the deep learning method.

Restricted Boltzmann machines (RBMs) \cite{Hinton1} are main deep learning
methods, they use energy-based learning models. The conditional probability
transformation for RBMs needs binary values \cite{Hinton1}. However, for
system identification the conditional probability distributions
$P(y|\mathbf{x})$ cannot be binary \cite{Jin}. In this paper, the RBMs are
modified such that the conditional probability distributions are continuous,
and the hidden weights can be trained by continuous input data.

Both fuzzy models and probability theory can represent and process uncertain
data effectively \cite{ChenF}. The dynamics and uncertainty in the data set in
many cases has probabilistic nature \cite{Gu}. The clustering methods
discussed above partition the data directly by calculating Euclidean
distances. These clusters do not include the distribution properties of
input/output data. They also do not scale well with large data sets due to the
quadratic computational complexity of calculating all the pair-wise distances
\cite{Li}. The clustering methods based on probability theory and models are
more powerful for big and uncertain data \cite{Gang}. On the other hand, we
use a restricted Boltzmann machine (RBM) to obtain the hidden features of the
joint vectorial space of the pairs input/output. The data obtained from the
RBM used for clustering are in the form of probability distributions. The
second contribution of this paper is that a probability based clustering
method is proposed to extract the fuzzy rules.

Including probability theory in fuzzy modeling can improve the stochastic
modeling capability \cite{Hu}. In \cite{Liu}, the probabilistic is added into
the fuzzy relation between the input space and the output space to handle the
effect of random noise and stochastic uncertainties. \cite{Waltman} introduces
probability distribution in the consequent part of the fuzzy rules improving
the fuzzy classifiers. In this paper, we introduce a probability parameter in
each fuzzy rule. This idea comes from the $Z$-number \cite{Zadeh1}, where a
probability measure is included into the fuzzy number to make the decision
fruitful based on human knowledge. The third contribution of this paper is we
apply probability parameters to classical fuzzy model and train these parameters.

\begin{figure}
[ptb]
\begin{center}
\includegraphics[
natheight=4.485800in,
natwidth=6.346900in,
height=3.4368in,
width=4.8542in
]%
{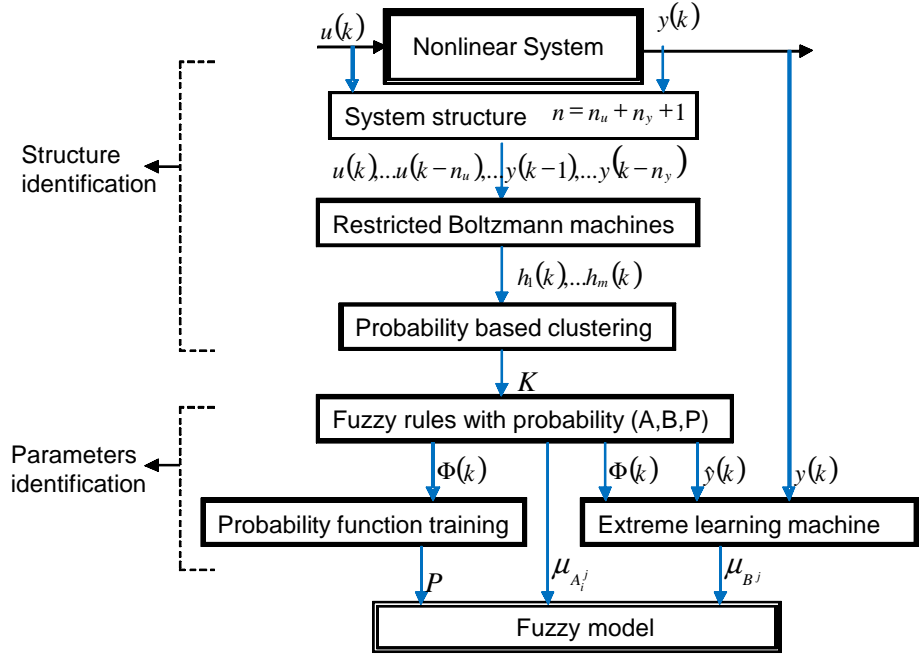}%
\caption{Data-driven fuzzy modeling}%
\label{f1}%
\end{center}
\end{figure}

Extreme learning machines \cite{Huang} and randomized algorithms
\cite{Schmidt} assign the hidden weights of a single hidden layer neural
network randomly and calculate the output weights with the pseudoinverse
approach (or least squares method). They have been successfully\ applied to
nonlinear system modeling \cite{Tapson}. \cite{Huang} shows that the
optimization of the hidden layer parameters does not improve the
generalization behavior significantly, while updating the output weights is
more effective. \cite{Igelnik} indicated that arbitrary assignment of the
hidden weights may lead to poor performances. In order to obtain good
approximation capability, in this paper we use RBMs and probability based
clustering to obtain the distributions of input/output features. We assign
these probability distributions as the hidden weights (the premise part of the
fuzzy rules). For the consequent part of the fuzzy rules (the output weights),
we use ELM to train them. Finally, we use an optimization method to reach
maximum probability\ measures in each fuzzy rule. The proposed data-driven
fuzzy modeling process is shown in Figure \ref{f1}.%

%TCIMACRO{\FRAME{ftbpFU}{4.8542in}{3.4368in}{0pt}{\Qcb{Data-driven fuzzy
%modeling}}{\Qlb{f1}}{f1.eps}{\special{ language "Scientific Word";
%type "GRAPHIC";  maintain-aspect-ratio TRUE;  display "USEDEF";
%valid_file "F";  width 4.8542in;  height 3.4368in;  depth 0pt;
%original-width 6.3469in;  original-height 4.4858in;  cropleft "0";
%filename 'f1.eps';file-properties "XNPEU";}} }%
%BeginExpansion
%EndExpansion
\bigskip

\section{Structure identification with deep learning and probability theory}

The following discrete-time nonlinear system is identified by our fuzzy
modeling method,%
\begin{equation}
y(k)=f\left[  \mathbf{x}\left(  k\right)  ,k\right]  \label{planta}%
\end{equation}
where%
\begin{equation}
\mathbf{x}\left(  k\right)  =[y\left(  k-1\right)  ,y\left(  k-2\right)
,\cdots y\left(  k-n_{y}\right)  ,u\left(  k\right)  ,u\left(  k-1\right)
,\cdots u\left(  k-n_{u}\right)  ]^{T} \label{X}%
\end{equation}
$f\left(  \cdot\right)  $ is an unknown nonlinear function, representing the
plant dynamics, $u\left(  k\right)  $ and $y\left(  k\right)  $ are the
measurable scalar input and output of the nonlinear plant, $n_{y}$ and $n_{u}$
correspond to the system order, $\mathbf{x}\left(  k\right)  \in\Re^{n}$ can
be regarded as a new input to the nonlinear function $f\left(  \cdot\right)
,$ $n=n_{y}+n_{u}+1.$ It is a NARMAX model \cite{Chen}.

The objective of the fuzzy modeling is to use the input and output data set
$\left[  y\left(  k\right)  ,\mathbf{x}\left(  k\right)  \right]  $ (or
$\left[  y\left(  k\right)  ,u\left(  k\right)  \right]  $) of the nonlinear
system (\ref{planta}), and construct a fuzzy model
\[
\hat{y}(k)=F\left[  \mathbf{x}\left(  k\right)  ,k\right]
\]
such that $\hat{y}(k)\rightarrow y(k),$ here $\hat{y}(k)$ is the output of the
fuzzy model $F\left[  \mathbf{\cdot}\right]  .$

This data-driven modeling scheme needs two basic processes: structure
identification and parameter identification. The structure identification is
to partition the input and the output data of the nonlinear system and extract
fuzzy rules.

As shown in \cite{Hinton}\cite{Bengio}, the restricted Boltzmann machine ()can
learn the probability distribution among the input data, and obtain their
hidden features. Moreover, a good RBM can help to improve the regression
accuracy significantly \cite{delaRosa}\cite{delaRosa1}\cite{delaRosa2}.

In this paper, we first use an RBM to transfer the input data to their feature
space, and obtain the hidden features of the input. The RBM transformation
allows us to model the system in the probability theory frame, such that the
model is not sensitivity to the noises and disturbances.

\subsection{Hidden feature learning with restricted Boltzmann machines}

The RBM can be regarded as a stochastic artificial neural network. It learns
the probability distribution of its input set. The input data to the RBM is
$\mathbf{x}\left(  k\right)  =\left[  x_{1}\cdots x_{n}\right]  ,$ the output
of the RBM is $\mathbf{\bar{h}}=\left[  \bar{h}_{1}\cdots\bar{h}_{m}\right]
.$ $n$ is the dimension of the input, $m$ is the dimension of the hidden
layer. If $i=1,...,m$ and $j=1,...,n,$ the $i-th$ hidden node and the $j-th$
visible node are%
\begin{equation}%
\begin{array}
[c]{c}%
p\left(  \bar{h}_{i}=1\mid\mathbf{x}\right)  =\phi\left[  V\mathbf{x}+b\right]
\\
p\left(  x_{j}=1\mid\mathbf{\bar{h}}\right)  =\phi\left[  V^{T}\mathbf{h}%
+c\right] \\
\bar{h}_{i}=\left\{
\begin{array}
[c]{cc}%
1 & a<p\left(  \bar{h}_{i}=1\mid\mathbf{x}\right) \\
0 & a\geq p\left(  \bar{h}_{i}=1\mid\mathbf{x}\right)
\end{array}
\right.
\end{array}
\label{rbm1}%
\end{equation}
where $\phi$ is the sigmoid function, $V$ is a weight matrix, $a$ is a
threshold, $b$ and $c$ are visible and hidden biases respectively.

We define the probability vector $\mathbf{h}$ as
\[
\mathbf{h}=\left[  p\left(  \bar{h}_{1}=1\mid\mathbf{x}\right)  \cdots
p\left(  \bar{h}_{m}=1\mid\mathbf{x}\right)  \right]  =\left[  h_{1}\cdots
h_{m}\right]
\]
The standard RBM model requires that both $\bar{h}_{i}$ and $x_{j}$ be binary
values. For nonlinear system identification, the visible units $\mathbf{x}$
cannot be binary values. We modify the standard RBM (\ref{rbm1}), such that it
is suitable for nonlinear system identification.

The objective of the training is to maximize the following probability
function%
\begin{equation}
p(\mathbf{x},\mathbf{\bar{h}})=\sum_{h}\frac{e^{-E(\mathbf{x},\mathbf{\bar{h}%
})}}{Z}\varpropto e^{-E(\mathbf{x},\mathbf{\bar{h}})} \label{pener}%
\end{equation}
where the energy function is $\ E(\mathbf{x},\mathbf{\bar{h}})=-\mathbf{\bar
{h}}^{T}V\mathbf{x}-b^{T}\mathbf{x}-c^{T}\mathbf{\bar{h},}$ the normalizing
factor $Z$ is called the partition function with $Z=\sum_{\mathbf{x}%
,\mathbf{h}}e^{-E\left(  \mathbf{x},\mathbf{h}\right)  }.$

In order to maximize $p(\mathbf{x},\mathbf{\bar{h}})$ with respect to the
weights $W$, the following free energy is defined%
\begin{equation}
\digamma(\mathbf{x})=-\log\sum_{\mathbf{\bar{h}}}e^{-E(\mathbf{x}%
,\mathbf{\bar{h}})} \label{fe}%
\end{equation}
If both $\overline{h}_{i}$ and $x_{j}$ are binary values, \textit{i. e.},
$\overline{h}_{i}$ and $x_{j}\mathbf{\in}\left\{  0,1\right\}  $, the
conditional probabilities are
\[
P(\mathbf{\bar{h}}|\mathbf{x})=\prod P(\overline{h}_{i}|\mathbf{x}),\quad
P(\mathbf{x}|\mathbf{\bar{h}})=\prod P(x_{j}|\mathbf{\bar{h}})
\]
However, for system identification the input $\mathbf{x}$ is continuous. To
handle this, $\mathbf{x}$ is first normalized in $[0,1].$ The conditional
probability for non-binary values in $[0,1]$ is calculated as follows.

The conditional probability for the $j$-th visible node is%
\[
P(x_{j}|\mathbf{\bar{h}})=\frac{e^{(V_{j}^{T}\overline{h}+c_{j})x_{j}}}%
{\int_{\widehat{x}_{j}}e^{(V_{j}^{T}\overline{h}+c_{j})\widehat{x}_{j}%
}d\widehat{x}_{j}}%
\]
When $x_{j}\in\lbrack0,1]$ the probability distribution with $a_{j}=V_{j}%
^{T}\overline{h}+c_{j}$,%
\begin{equation}
P(x_{j}|\mathbf{\bar{h}})=\frac{a_{j}e^{a_{j}x_{j}}}{e^{a_{j}}-1} \label{c2}%
\end{equation}
The cumulative conditional probability from where a sampling process can be
made is computed by
\begin{equation}
P_{C}(x_{j}|\mathbf{\bar{h}})=\frac{e^{a_{j}x_{j}}-1}{e^{a_{j}}-1}
\label{cpd2}%
\end{equation}
Finally the expected value of the distribution is
\begin{equation}
E[x_{j}]=\frac{1}{1-e^{-a_{j}}}-\frac{1}{a_{j}} \label{ev2}%
\end{equation}
\bigskip We use the data set, $\mathbf{x}\left(  k\right)  \in D_{1}$
(training set)$,$ $k=1\cdots M,$ to train the RBM (\ref{rbm1}). If we define
the parameters as $\Theta=\left[  V,b,c\right]  ,$ the gradient descent method
is%
\begin{equation}
\theta\left(  k+1\right)  =\theta\left(  k\right)  -\eta\frac{\partial\left[
-\log P\left(  x\right)  \right]  }{\partial\theta\left(  k\right)  },\quad
k=1\cdots M \label{up3}%
\end{equation}
where $\eta>0$ is the learning rate. This stochastic gradient descent
algorithm can minimize the function $\left[  -\log P\left(  \mathbf{x}\right)
\right]  $. The log-likelihood gradient of $P\left(  x\right)  $ with respect
to $\theta\in\Theta$ is%
\[
\frac{\partial\log P\left(  \mathbf{x}\right)  }{\partial\theta\left(
k\right)  }=\sum_{\widehat{\mathbf{x}}}P\left(  \widehat{\mathbf{x}}\right)
\frac{\partial\digamma(\widehat{\mathbf{x}})}{\partial\theta\left(  k\right)
}-\frac{\partial\digamma(\mathbf{x})}{\partial\theta\left(  k\right)  }%
\]
where $\sum_{\widehat{\mathbf{x}}}$ indicates a sum along the entire sampling
space of $x$. Here $\sum_{\widehat{\mathbf{x}}}P\left(  \widehat{\mathbf{x}%
}\right)  \frac{\partial\digamma(\widehat{\mathbf{x}})}{\partial\theta\left(
k\right)  }$ is estimated by the contrastive divergence approximation (CD)
\cite{Hinton},
\[
\sum_{\widehat{\mathbf{x}}}P\left(  \widehat{\mathbf{x}}\right)
\frac{\partial\digamma(\widehat{\mathbf{x}})}{\partial\theta\left(  k\right)
}\thickapprox\frac{\partial\digamma(\widetilde{\mathbf{x}})}{\partial\theta}%
\]
This approach uses the Gibbs sampling to create an estimate of the input
expectation $\widetilde{\mathbf{x}}$. Usually $\widetilde{\mathbf{x}}$ is
estimated by one step-Gibbs sampling, which provides a good trade-off between
speed and accuracy \cite{Bengio}.

After the RBM (\ref{rbm1}) is trained, the parameters $\Theta$ are fixed. Then
we use another data set, $\mathbf{x}\left(  k\right)  \in D_{2}$, $k=1\cdots
N,$ to do the data-driven fuzzy modeling. $N$ is the number of training
examples. Now the RBM transforms the input data to their hidden feature space.

Because the features of the input data are in the form of probability
distributions, we use the following probability based clustering method to
obtain the fuzzy rules.

\subsection{Probability based clustering}

The input data $\mathbf{x}\left(  k\right)  \in D_{2}$ are mapped to the
hidden features $H=\{\mathbf{h}(k)\}_{k=1}^{N}$ by the trained RBM
(\ref{rbm1}). We assume each sample $\mathbf{h}(k)$ belongs to a specific
cluster whose labels are given by $L=\{l(k)\}_{k=1}^{N}$, $l(k)\in
\{1,...,K\},$ where $K$ is the number of clusters. The object of the
probability based clustering is to find the correlation between the input
instances and their respective cluster parameters. The higher correlation
between an instance and a cluster, the more possible it will be assigned to
that cluster. We use the following objective function, which is similar as
\cite{Gang},%
\begin{equation}
P(L,\{\delta_{j}\}_{j=1}^{K}|H)\varpropto p(L)\left[  \prod_{k=1}%
^{N}p(\mathbf{h}(k)|\delta_{l(k)})\right]  \prod_{j=1}^{K}p(\delta_{j})
\label{veroClus}%
\end{equation}
where $p(L)$ is the marginal clustering distribution probability, $\delta_{j}$
are the clustering model parameters, $p(\mathbf{h}(k)|\delta_{l(k)})$ is the
likelihood of the hidden code $\mathbf{h}(k),$ $\delta_{l(k)}$ is the cluster
parameter$,$ $p(\delta_{j})$ is the Gaussian prior for all $\delta_{j}$ with
$j=1...K.$

The parameters $\{\delta_{j}\}_{j=1}^{K}$ are estimated by the\ following
Gibbs sampling with respect to the label $l(k)$ and hidden feature
$\mathbf{h}(k).$ Given the set of codes $H=\{\mathbf{h}(k)\}_{k=1}^{N}$ and
its cluster labels $L,$ the Gibbs sampling allow us to obtain samples from the
conditional probability distribution while keeping other variables fixed. So
for each label $l(k)$, the conditional posterior is%
\begin{equation}
p\left[  l(k)=j|l(-k),\mathbf{h}(k),\{\delta_{j}\}_{j=1}^{K},\alpha
,\psi,\lambda\right]  \varpropto p\left[  l(k)=j|l(-k),\alpha,\psi\right]
p\left[  \mathbf{h}(k)|\delta_{j}\right]  \label{GibbsSam}%
\end{equation}
where $l(-k)$ denotes all\ other indices but $k.$

$p\left[  l(k)=j|l(-k),\alpha\right]  $ is determined by a Chinese restaurant
process with concentration parameter $\alpha$ and discount parameter $\psi$.
The probability of each cluster given by the Chinese restaurant process is
calculated as follows: at time $k+1$ suppose that we have $K$ different
clusters, then $\mathbf{h}(k)$ would be assigned at an empty new cluster
$G_{K+1}$ with probability $\frac{\psi+K\alpha}{k+\psi}$. For an existing
cluster $G_{j}$ with $n_{j}$ existing elements, the probability is
$\frac{n_{j}-\alpha}{k+\psi}.$

$p(\mathbf{h}(k)|\delta_{j})$ is the likelihood for the current instance $k$
and $\mathbf{h}(k)$ in its cluster. It is directly proportional to the
correlation between $\mathbf{h}(k)$ and $\delta_{j}$. It can be calculated as
$\mathbf{h}(k)^{T}\delta_{j}.$ Taking into account the weight penalization
$\lambda\left\Vert \delta_{j}\right\Vert ^{2}$, it can also be calculated as
\begin{equation}
p(\mathbf{h}(k)|\delta_{j})\varpropto\exp(\mathbf{h}(k)^{T}\delta_{j}%
-\lambda\left\Vert \delta_{j}\right\Vert ^{2}) \label{PRetiq}%
\end{equation}
where $\lambda$ is a penalization constant to control the weights size,
$\lambda\left\Vert \delta_{j}\right\Vert ^{2}$ represents the maximum margin
to separate clusters \cite{Gang}.

(\ref{PRetiq}) is regarded as a set of exponential functions, which have
similar statistics properties. Substituting the assumption (\ref{GibbsSam})
into (\ref{PRetiq}),%
\begin{equation}
p\left[  l(k)=j|l(-k),\mathbf{h}(k),\{\delta_{j}\}_{j=1}^{K},\alpha
,\lambda\right]  \varpropto p\left[  l(k)=j|l(-k),\alpha\right]
\exp(\mathbf{h}(k)^{T}\delta_{j}-\lambda\left\Vert \delta_{j}\right\Vert ^{2})
\label{GibbsSam2}%
\end{equation}

A lager correlation between $\mathbf{h}(k)$ and $\delta_{j}$ indicates a
higher probability that $\mathbf{h}(k)$ belongs to cluster $G_{j}$. If the
probability is less than a probability threshold, a new virtual cluster
$G_{K+1}$ with random parameters $\delta_{K+1}$ is generated, $K=K+1.$

$\mathbf{h}(k)$ is assigned into this new cluster. The probability of a new
cluster is calculated by the Chinese restaurant process. The correlation is
calculated by (\ref{GibbsSam2})$.$ $\delta_{K+1}$ is drawn from a
multi-variate $t$-distribution.

So the clustering object is to maximize (\ref{veroClus}) as%
\begin{equation}
\max\left\{  p(L)\left[  \prod_{k=1}^{N}p(\mathbf{h}(k)|\delta_{l(k)})\right]
\prod_{j=1}^{K}p(\delta_{j})\right\}  \label{ob}%
\end{equation}
The probabilities $p(\delta_{j})$ is calculated by the following maximum
margin learning rule. The maximum margin learning rule uses the passive
aggressive algorithm (PA) \cite{Crammer} to update the cluster parameters. At
time $k,$ the label $l(k)$ is determined by the Gibbs sampling process
described in (\ref{GibbsSam2}).

We concatenate the cluster parameters $\{\delta_{j}\}_{j=1}^{K}$ as a vector
$\Delta=[\delta_{1},...,\delta_{K}],$ or $\Delta^{l(k)}=\delta_{l(k)}$. If we
define the concatenating vector $\Phi\left[  \mathbf{h}(k),l(k)\right]  $
where the $l(k)-th$ element is set to be $\mathbf{h}(k),$ while the others are
set to be vectors $0$ we calculate at time $k$ the vector $\Delta(k)$ margin
as
\begin{equation}
M\left[  \Delta(k);(\mathbf{h}(k),l(k))\right]  =\Delta(t)\cdot\Phi\left[
\mathbf{h}(k),l(k)\right]  -\Delta(k)\cdot\Phi\left[  \mathbf{h}%
(k),\widehat{l}(k)\right]  \label{mm}%
\end{equation}
where $\widehat{l}(k)$ is the prediction label from the model and
$\mathbf{h}(k)$,%
\begin{equation}
\widehat{l}(k)=\arg\max_{j}\mathbf{h}(k)^{T}\delta_{j} \label{correlationMax}%
\end{equation}

The updating process is designed to optimize the following objective function%
\begin{equation}%
\begin{array}
[c]{cc}
& \Delta(k+1)=\arg\min_{\Delta}\frac{1}{2}\left\Vert \Delta-\Delta
(k)\right\Vert ^{2}+C\xi\\
\text{Subject:} & l_{2}\left[  \Delta;(\mathbf{h}(k),l(k))\right]  \leq\xi
\end{array}
\label{minprob}%
\end{equation}
where $C>0$ is a penalty constant, $\xi$ is the threshold of the hinge-loss
function, $l_{2}\left[  \cdot\right]  $ is the hinge-loss function defined by
\begin{equation}
l_{2}\left[  \Delta(k);(\mathbf{h}(k),l(k))\right]  =\left\{
\begin{array}
[c]{cc}%
0 & \text{if }M\left[  \Delta(k);(\mathbf{h}(k),l(k))\right]  \geq1\\
1-M\left[  \Delta(k);(\mathbf{h}(k),l(k))\right]  & \text{otherwise}%
\end{array}
\right.  \label{hinge}%
\end{equation}
where $M\left[  \cdot\right]  $ is the margin function (\ref{mm})$.$

Using the passive aggressive algorithm \cite{Crammer}, the parameters are
updated as
\begin{equation}%
\begin{array}
[c]{c}%
\Delta^{l(k)}(k+1)=\Delta^{l(k)}(k)+\tau(k)\mathbf{h}(k)\\
\Delta^{\widehat{l}(k)}(k+1)=\Delta^{\widehat{l}(k)}(k)-\tau(k)\mathbf{h}(k)
\end{array}
\label{updatez}%
\end{equation}
where $\tau(k)=\min\{C,\frac{hl\left[  \Delta(k);(\mathbf{h}(k),l(k))\right]
}{\left\Vert \mathbf{h}(k)\right\Vert ^{2}}\}$.

For each iteration $k$, $\delta_{k}$ is estimated by (\ref{updatez}),
(\ref{hinge}), and (\ref{mm}), such that the maximum margin archives. This
probability based clustering is similar as the nonparametric maximum margin
clustering \cite{Gang}. However, the data of this paper are time series and
the algorithm of this paper can be applied on-line.

\subsection{Fuzzy rules extraction with probability theory}

After the probability based clustering, we have $K$ different clusters
$G_{j},$ $j=1\cdots K.$ We assign one fuzzy rule for each cluster $G_{j}$ as%
\begin{equation}
\text{R}^{j}\text{: IF }h_{1}\left(  k\right)  \text{ is }A_{1\text{ }}%
^{j}\text{and }h_{2}\left(  k\right)  \text{ is }A_{2}^{j}\text{ and }%
\cdots\text{ }h_{m}\left(  k\right)  \text{ is }A_{m}^{j}\text{ THEN }y\left(
k\right)  \text{ is }B^{j} \label{fuzru}%
\end{equation}
where $A_{1\text{ }}^{j},\cdots A_{m}^{j}$ and $B^{j}$ are standard fuzzy
sets, they are represented by the following Gaussian membership functions%
\begin{equation}
\mu_{A_{i}^{j}}\left[  h_{i}\left(  k\right)  \right]  =\exp\left(
-\frac{\left[  h_{i}\left(  k\right)  -c_{ji}\right]  ^{2}}{\sigma_{ji}^{2}%
}\right)  \label{Gau}%
\end{equation}
where $k=1\cdots N,$ $i=1\cdots m,$ $j=1\cdots K.$

By using product inference, center-average and singleton fuzzifier, the output
of the fuzzy system is expressed as \cite{Wang}
\begin{equation}
\hat{y}\left(  k\right)  =\left(  \sum\limits_{j=1}^{K}w_{j}\left[
\prod\limits_{i=1}^{n}\mu_{A_{i}^{j}}\right]  \right)  /\left(  \sum
\limits_{j=1}^{K}\left[  \prod\limits_{i=1}^{n}\mu_{A_{i}^{j}}\right]
\right)  \label{fuz1}%
\end{equation}
where $w_{j}$ is the point at which $\mu_{B^{j}}=1$. If we define $\phi
_{j}=\prod\limits_{i=1}^{n}\mu_{A_{i}^{j}}/\sum\limits_{j=1}^{K}%
\prod\limits_{i=1}^{n}\mu_{A_{i}^{j}},$ (\ref{fuz1}) can be expressed in
matrix form
\begin{equation}
\hat{y}\left(  k\right)  =\mathbf{W}\left(  k\right)  \Phi\left[
\mathbf{h}\left(  k\right)  \right]  \label{fuzneu1}%
\end{equation}
with parameters $\mathbf{W}\left(  k\right)  =\left[  w_{1}\cdots
w_{K}\right]  $ and data vector $\Phi\left[  \mathbf{h}\left(  k\right)
\right]  =\left[  \phi_{1}\cdots\phi_{K}\right]  ^{T}.$

\begin{figure}
[ptb]
\begin{center}
\includegraphics[
natheight=6.201600in,
natwidth=9.367600in,
height=3.4835in,
width=5.2174in
]%
{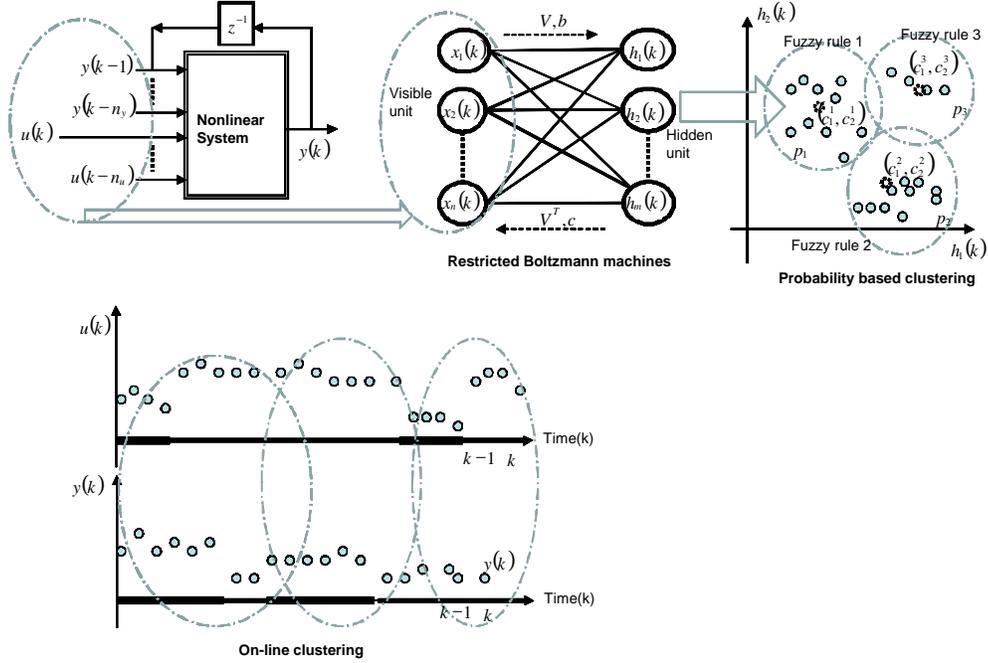}%
\caption{Fuzzy rules extraction with the on-line culstering and the
probability based clustering}%
\label{f2}%
\end{center}
\end{figure}

From the restricted Boltzmann machine, we obtain the hidden features
$h_{i}\left(  k\right)  $ and their dimension $m.$ From the probability based
clustering, we obtain the fuzzy rule number $K$ and the data distributions. So
the structure of the fuzzy model is ready. The fuzzy rules extraction with the
on-line clustering and the probability based clustering is shown in Figure
\ref{f2}.%

%TCIMACRO{\FRAME{ftbpFU}{5.2174in}{3.4835in}{0pt}{\Qcb{Fuzzy rules extraction
%with the on-line culstering and the probability based clustering}}{\Qlb{f2}%
%}{f2.eps}{\special{ language "Scientific Word";  type "GRAPHIC";
%display "USEDEF";  valid_file "F";  width 5.2174in;  height 3.4835in;
%depth 0pt;  original-width 9.3676in;  original-height 6.2016in;
%cropleft "0";  croptop "1";  cropright "1";  cropbottom "0";
%filename 'f2.eps';file-properties "XNPEU";}} }%
%BeginExpansion
%EndExpansion

The probability based clustering not only gives the distribution of the data
$h_{i}\left(  k\right)  $, but also provides the relations of the data in
probability forms. The fuzzy rule (\ref{fuzru}) only represents the data
distribution. In order to include the flexibility of this probability relation
in the data, we assign probability factors $p_{j,i}$ into each rule
(\ref{fuzru}) as%
\begin{equation}%
\begin{array}
[c]{c}%
\text{R}^{j}\text{: IF }h_{1}\left(  k\right)  \text{ is }A_{1\text{ }}%
^{j}\text{and }h_{2}\left(  k\right)  \text{ is }A_{2}^{j}\text{ and }%
\cdots\text{ }h_{m}\left(  k\right)  \text{ is }A_{n}^{j}\text{ THEN}\\%
\begin{array}
[c]{l}%
y\left(  k\right)  \text{ is }B^{1}\text{ with prob. }p_{j,1}\text{ and}\\
y\left(  k\right)  \text{ is }B^{2}\text{ with prob. }p_{j,2}\text{ and}\\
...\\
y\left(  k\right)  \text{ is }B^{K}\text{ with prob. }p_{j,K}\text{ }%
\end{array}
\end{array}
\label{frp}%
\end{equation}
where $p_{j,i}\geq0$, $\sum_{i=1}^{K}p_{j,i}=1$ with $i,j=1,...,K$. This means
the consequent $y\left(  k\right)  $ is established in the probability given
by $p_{j,i}.$ So the fuzzy set of the consequent, $B^{j},$ should satisfy%
\begin{equation}
p(B^{j}|\mathbf{h}(k))=\sum_{i=1}^{K}\phi_{i}\left[  \mathbf{h}\left(
k\right)  \right]  p_{i,j} \label{consX}%
\end{equation}

\section{Data-Driven Fuzzy Modeling}

The fuzzy model of the probability based fuzzy rules is not longer
(\ref{fuzneu1}). We use the following process to extract the fuzzy model from
the feature space $\mathbf{h}(k).$ $\phi_{j}\left[  \mathbf{h}\left(
k\right)  \right]  $ in (\ref{consX}) can be regarded as a normalized
vectorial membership function of $\mathbf{h}\left(  k\right)  $ to the fuzzy
sets $A_{1}^{j},A_{2}^{j},\cdots,A_{m}^{j}$, $p(y|\mathbf{h}(k))$ is
calculated by%
\begin{equation}
p(y|\mathbf{h}(k))=\sum_{j=1}^{K}p(y|B^{j})p(B^{j}|\mathbf{h}(k))
\label{YGivX}%
\end{equation}
where $p(y|B^{j})$ is estimated as%
\begin{equation}
p(y|B^{j})=\frac{\mu_{B^{j}}(y)}{\int\mu_{Bj}(y)dy} \label{pyb}%
\end{equation}
This is a probability measurement for the membership function $\mu_{B^{j}}.$
The output of the probability based fuzzy model is%
\begin{equation}
\widehat{y}\left(  k\right)  =E(y|\mathbf{h}(k))=\int yp(y|\mathbf{h}%
(k))dy=\sum_{j=1}^{K}p(B^{j}|\mathbf{h}(k))E(y|B^{j}) \label{pfuzzy}%
\end{equation}
where $E(y|B^{j})=\frac{\int y\mu_{B^{j}}(y)dy}{\int\mu_{Bj}(y)dy}.$ The last
term is just the centroid of the fuzzy set $B^{j}.$

Compared with the standard fuzzy model (\ref{fuzneu1}), where $w_{j}$ is the
point at which $\mu_{B^{j}}=1$, (\ref{pfuzzy}) can be formed as
\begin{equation}
\widehat{y}\left(  k\right)  =\sum_{j=1}^{K}w_{j}p(B^{j}|\mathbf{h}%
(k))=\sum_{j=1}^{K}\sum_{i=1}^{K}\phi_{i}\left[  \mathbf{h}\left(  k\right)
\right]  p_{i,j}w_{j} \label{pfuzzy1}%
\end{equation}
or
\begin{equation}
\hat{y}\left(  k\right)  =\mathbf{W}\left(  k\right)  \Phi\left[
\mathbf{h}\left(  k\right)  \right]  \label{pfuzzy2}%
\end{equation}
where the parameter $\mathbf{W}\left(  k\right)  =\left[  w_{1}\cdots
w_{K}\right]  $ and the data vector%
\[
\Phi\left[  \mathbf{h}\left(  k\right)  \right]  =\left[  \sum_{i=1}^{K}%
\phi_{i}\left[  \mathbf{h}\left(  k\right)  \right]  p_{i,1}|\cdots|\sum
_{i=1}^{K}\phi_{i}\left[  \mathbf{h}\left(  k\right)  \right]  p_{i,K}\right]
^{T}%
\]

\subsection{Extreme learning machine for membership functions training}

For the probability based fuzzy model (\ref{pfuzzy2}), $\Phi\left[
\mathbf{h}\left(  k\right)  \right]  $ is determined by the restricted
Boltzmann machine and probability based clustering as we present below.
(\ref{pfuzzy2}) is a linear-in-parameter system, the parameter $\mathbf{W}%
\left(  k\right)  $ may be singular and/or be not square, the solution can be
solved by the Moore-Penrose generalized inverse, which is defined as follows.

\begin{definition}
The matrix $A^{+}\in\Re^{n\times m}$ is the Moore-Penrose generalized inverse
of $A\in\Re^{m\times n}$ if%
\begin{equation}
AA^{+}A=A,\text{\quad}A^{+}AA^{+}=A^{+},\quad\left(  AA^{+}\right)
^{T}=AA^{+},\quad\left(  A^{+}A\right)  ^{T}=A^{+}A \label{genInv}%
\end{equation}

\end{definition}

In particular, when $A$ has full column rank,
\begin{equation}
A^{+}=\left(  A^{T}A\right)  ^{-1}A^{T} \label{ap}%
\end{equation}
When $A$ has full row rank
\begin{equation}
A^{+}=A^{T}\left(  AA^{T}\right)  ^{-1} \label{ap1}%
\end{equation}

\begin{definition}
$x_{0}\in$ $\Re^{n}$ is said to be a minimum norm least-squares solution of
the linear system $y=Ax$ if
\begin{equation}
\left\Vert x_{0}\right\Vert \leq\left\Vert x\right\Vert ,\quad\forall
x\in\left\{  x:\left\Vert Ax-y\right\Vert \leq\left\Vert Az-y\right\Vert
,\forall z\in\Re^{n}\right\}  \label{minNorm}%
\end{equation}
where $y\in\Re^{m}.$
\end{definition}

For a linear system $\hat{y}\left(  k\right)  =\mathbf{W}\Phi$, $\mathbf{W}%
_{0}$ is a least-squares solution if
\begin{equation}
\left\Vert \mathbf{W}_{0}\Phi-y\left(  k\right)  \right\Vert =\underset
{W}{\min}\left\Vert \mathbf{W}\Phi-y\left(  k\right)  \right\Vert \label{ls}%
\end{equation}
where $\left\Vert \cdot\right\Vert $ is a norm in Euclidean space. If $By$ is
a minimum norm least-squares solution of the linear system $\hat{y}%
=\mathbf{W}\Phi,$ then it is necessary and sufficient that $B=\Phi^{+}$. Here
$\Phi^{+}$ is the Moore-Penrose generalized inverse of matrix $\Phi$, which is
defined in (\ref{genInv}).

For our fuzzy model, the goal of the training algorithm is to find the
parameter $\mathbf{W}\left(  k\right)  $ such that the following cost function
is minimized
\begin{equation}
J=\sum_{k}\left\Vert y\left(  k\right)  -\widehat{y}\left(  k\right)
\right\Vert ^{2} \label{ind}%
\end{equation}
The training data are $y\left(  k\right)  $ and $\Phi\left[  \mathbf{h}\left(
k\right)  \right]  $, $k=1,2\cdots N$, $N$ is the total training data number.

Considering the entire training set,
\begin{equation}
\hat{Y}=\left[
\begin{tabular}
[c]{llll}%
$\hat{y}\left(  1\right)  $ & $\hat{y}\left(  2\right)  $ & $\cdots$ &
$\hat{y}\left(  N\right)  $%
\end{tabular}
\right]  =\left[
\begin{tabular}
[c]{llll}%
$\mathbf{W}\Phi\left(  1\right)  $ & $\mathbf{W}\Phi\left(  2\right)  $ &
$\cdots$ & $\mathbf{W}\Phi\left(  N\right)  $%
\end{tabular}
\right]  =\mathbf{W}\Psi\label{pseu1}%
\end{equation}
where $\Psi=\left[  \Phi\left(  1\right)  ,\Phi\left(  2\right)  ,\cdots
,\Phi\left(  N\right)  \right]  .$ Or in another form:%
\begin{align}
Y  &  =\left[
\begin{tabular}
[c]{llll}%
$y\left(  1\right)  $ & $y\left(  2\right)  $ & $\cdots$ & $y\left(  N\right)
$%
\end{tabular}
\right]  =\left[
\begin{tabular}
[c]{llll}%
$\mathbf{W}\Phi\left(  1\right)  +e(1)$ & $\mathbf{W}\Phi\left(  2\right)
+e(2)$ & $\cdots$ & $\mathbf{W}\Phi\left(  N\right)  +e(N)$%
\end{tabular}
\ \ \ \ \ \right] \nonumber\\
Y  &  =\mathbf{W}\Psi+E \label{ps2}%
\end{align}
where $e\left(  k\right)  $ is the modeling error $e(k)=y\left(  k\right)
-\widehat{y}\left(  k\right)  $, and $E=\left[  e\left(  1\right)  ,e\left(
2\right)  ,\cdots,e\left(  N\right)  \right]  .$ To obtain $\min
\limits_{\beta}J,$ \ we need $\frac{\partial J}{\partial\mathbf{W}}=0.$ From
(\ref{ap1})
\begin{equation}
\mathbf{W}^{\ast}=Y\Psi^{T}\left(  \Psi\Psi^{T}\right)  ^{-1}=Y\Psi^{+}
\label{lss}%
\end{equation}
So $\mathbf{W}^{\ast}$ can minimize the index $J$ in (\ref{ind}).

Since $\mathbf{W}^{\ast}$\ is one of the least-squares solution of the system
$Y=\mathbf{W}\Psi+E$, it reaches the smallest approximation error on the
training data set, and it is unique. The solution $\mathbf{W}^{\ast}$ has the
smallest norm for a least-squares solution of $Y=\mathbf{W}\Psi.$
\cite{Schmidt} shows that for feedforward networks, small norm of the weights
is more important than the node number to obtain small generalization error.

The extreme learning machine \cite{Huang} and the randomized algorithm
\cite{Igelnik} require arbitrary assignment for the hidden weights. Although
random weights in the hidden layers are better than backpropagation training
in many cases, sometimes random weights may lead to poor performances
\cite{Igelnik}. The restricted Boltzmann machine and the probability based
clustering provide possible selection manners of hidden weights with the
distribution of the input data. The distributions of the random hidden weights
are defined in advance to improve the modeling accuracy.

For the fuzzy model, the premise membership functions $A_{1\text{ }}%
^{j},\cdots A_{m}^{j}$ are given by the probability based clustering.
$A_{i\text{ }}^{j}$ is in the form of Gaussian function (\ref{Gau}). Its two
parameters $c_{ji}$ and $\sigma_{ji}$ are determined as:

\begin{itemize}
\item The terms $c_{ji}$ are selected as equal as the center of each cluster

\item The parameters $\sigma_{ji}$ are assigned randomly in$\left(
0,1\right)  $
\end{itemize}

As we do not have the values of $p_{j,i}$ we cannot calculate $W$. We set the
parameters $p_{j,i}=1$ for $i=j$ and $p_{j,i}=0$ for $i\neq j$ which reduces
the probabilistic model (\ref{frp}) into the model (\ref{fuzru}). With this
consideration we can compute $W.$ The next sub-section shows how to estimate
the probability parameters $p_{j,i}$.

\subsection{Probability functions training}

The object of training the probabilities $p_{j,i}$ of each fuzzy rule
(\ref{frp}) is to maximize the likelihood of the desired output with respect
to its input. From (\ref{consX}) and (\ref{YGivX}), the parameters $p_{i,j}$
satisfy%
\begin{equation}
p(y|\mathbf{h}\left(  k\right)  )=\sum_{j=1}^{K}p(y|B^{j})\sum_{i=1}^{K}%
\phi_{i}\left[  \mathbf{h}\left(  k\right)  \right]  p_{i,j} \label{probYest}%
\end{equation}
Because $p_{i,K}=1-\sum_{j=1}^{K-1}p_{i,j},$
\begin{equation}
p(y|\mathbf{h}\left(  k\right)  )=\sum_{j=1}^{K-1}p(y|B^{j})\sum_{i=1}^{K}%
\phi_{i}\left[  \mathbf{h}\left(  k\right)  \right]  p_{i,j}+p(y|B^{K}%
)\sum_{i=1}^{K}\phi_{i}\left[  \mathbf{h}\left(  k\right)  \right]  \left(
1-\sum_{j=1}^{K-1}p_{i,j}\right)  \label{probYDes}%
\end{equation}
Then the global log-likelihood function of the training set $D$ such that
$\left\{  \mathbf{h}\left(  k\right)  ,y(k)\right\}  \in D$ is%
\begin{equation}%
\begin{array}
[c]{l}%
%TCIMACRO{\tciLaplace}%
%BeginExpansion
\mathcal{L}%
%EndExpansion
(D,P)=\log\left(  \prod_{k=1}^{N}p(y(k)|\mathbf{h}\left(  k\right)  )\right)
=\sum_{k=1}^{N}\log p(y(k)|\mathbf{h}\left(  k\right)  )\\
=\sum_{k=1}^{N}\log\left[
\begin{array}
[c]{c}%
\sum_{j=1}^{K-1}p(y|B^{j})\sum_{i=1}^{K}\phi_{i}\left[  \mathbf{h}\left(
k\right)  \right]  p_{i,j}+\\
p(y|B^{K})\sum_{i=1}^{K}\phi_{i}\left[  \mathbf{h}\left(  k\right)  \right]
\left(  1-\sum_{j=1}^{K-1}p_{i,j}\right)
\end{array}
\right]
\end{array}
\end{equation}
where $P$ is a $K\times K$ dimension matrix which contains the probability
parameters $p_{j,i},$
\begin{equation}
P=\left[
\begin{array}
[c]{ccc}%
p_{1,1} & \cdots & p_{1,K}\\
\vdots & \ddots & \vdots\\
p_{K,1} & \cdots & p_{K,K}%
\end{array}
\right]  \label{probabilidades}%
\end{equation}
The fuzzy set $B^{j}$ has the form of a Gaussian function (\ref{Gau}) with
$c_{j}=w_{j}$,%
\[
\mu_{B^{j}}\left(  y(k)\right)  =\exp\left(  -\frac{\left(  y(k)-c_{j}\right)
^{2}}{\sigma_{B^{j}}^{2}}\right)
\]
By using $\int\mu_{B^{j}}(y)dy=\sqrt{\pi}\sigma_{B^{j}}$, we can evaluate
$p(y(k)|B^{i}).$

In order to obtain $P,$ we need to solve the following minimization problem%
\begin{equation}
\left\{
\begin{array}
[c]{cc}
& \min_{P}\left\{  -%
%TCIMACRO{\tciLaplace}%
%BeginExpansion
\mathcal{L}%
%EndExpansion
(D,P)\right\} \\
\text{Subject} & p_{i,j}>0\forall\text{ }i,j\text{ and }\sum_{j=1}%
^{K-1}p_{i,j}\leq1
\end{array}
\right.  \label{MinProbsDif}%
\end{equation}
Here we do not use the last column $p_{i,K}$ of $P,$ because it is calculated
as a consequence of the rest of the values of $P.$

The minimization (\ref{MinProbsDif}) can be formed into the following linear
programming program as%
\begin{equation}
\left\{
\begin{array}
[c]{cc}
& \min_{P_{v}}-%
%TCIMACRO{\tciLaplace}%
%BeginExpansion
\mathcal{L}%
%EndExpansion
(D,P_{v})\\
\text{Subject} & AP_{v}\leq b\text{ and }l_{b}\leq P_{v}%
\end{array}
\right.  \label{MinProbModif}%
\end{equation}
where%
\[%
\begin{array}
[c]{c}%
P_{v}=[p_{1},\cdots,p_{K-1}%
%TCIMACRO{\U{a6}}%
%BeginExpansion
\vert
%EndExpansion
\cdots%
%TCIMACRO{\U{a6}}%
%BeginExpansion
\vert
%EndExpansion
p_{K},\cdots,p_{K,K-1}]^{T}\\
A=\left[
\begin{array}
[c]{cccc}%
\overrightarrow{1} & \overrightarrow{0} & \cdots & \overrightarrow{0}\\
\overrightarrow{0} & \overrightarrow{1} & \cdots & \overrightarrow{0}\\
\vdots & \vdots & \ddots & \vdots\\
\overrightarrow{0} & \overrightarrow{0} & \cdots & \overrightarrow{1}%
\end{array}
\right]
\end{array}
\]
$\overrightarrow{0}$ ,$\overrightarrow{1}\in%
%TCIMACRO{\U{211d} }%
%BeginExpansion
\mathbb{R}
%EndExpansion
^{K-1}$ are row vectors with $\overrightarrow{0}=[0...0]$ and $\overrightarrow
{1}=[1...1]$, $l_{b},b\in%
%TCIMACRO{\U{211d} }%
%BeginExpansion
\mathbb{R}
%EndExpansion
^{K-1}$ such that $b=[1...1]^{T}$ and $l_{b}=[0...0]^{T}.$ The minimization
problem of (\ref{MinProbsDif}) is solved by a standard linear programming
toolbox of Matlab.

\section{Comparisons with other fuzzy modeling methods}

In this section, we use two benchmark examples to show the effectiveness of
our data-driven fuzzy modeling method which combines the restricted Boltzmann
machines, the probability based clustering, and probability fuzzy rules.

\subsection{Gas furnace modeling}

The first example is the famous gas furnace data from the textbook \cite{Box}.
In this data set, the air and methane are mixed to generate mixture gas which
contains the carbon dioxide. The methane is regarded as\ input, $u(k),$ while
the CO$_{2}$ is the output $y(k)$. There are $296$ successive pairs of
observations $[u(k),y(k)],$ which are measured from the continuous records in
$9$ seconds. A general model is%
\[
y(k)=f[y(k-1),\ldots,y(k-n_{y}),u(k),\ldots,u(k-n_{u})]=f\left[
\mathbf{x}\left(  k\right)  \right]
\]
where $n_{y}$ and $n_{u}$ are the regression delays for the input and the output.

Here we use the random search method \cite{Collobert}\cite{Bergstra} to decide
the best $n_{y}$ and $n_{u}.$ The regression delays are assumed in the
interval $[1,10],$ the training data are $200$ examples while the rest are
used for validation. Finally, we have $n_{y}=4,$ $n_{u}=5.$

The data set is first normalized for comparison purposes. In this paper, the
data-driven fuzzy modeling has the following four steps:

\begin{enumerate}
\item Features extraction. The normalized input data are sent to an RBM: The
contrastive divergence uses $1$-step Gibbs sampling and $10$ training epochs,
the learning rate is $\eta=0.2$. After the training, the parameters of the RBM
$V$ and $b$ are then used to compute the hidden representation of the model
($\mathbf{h}$). The number of hidden units is chosen as $n_{y}+n_{u}+1,$ such
that the hidden and the visible unit numbers are the same.

\item Clustering. After the features are extracted by the RBM, we used the
probability based clustering. The hyper parameters are chosen as $\alpha=0.8$,
$\psi=10,$ $\lambda=5$ and $C=0.001.$ Here $\alpha$ and $\psi$ determine the
probabilities which\ are obtained by the Chinese restaurant process. $\alpha$
is close to $1.$ When $\psi$ increases, the number of clusters $K$ also grows.
The penalization parameter $\lambda$ decreases the probability of the cluster,
while keeps $\left\Vert \delta_{j}\right\Vert ^{2}$ low . In our experiments,
the probability based clustering divided the data set $\mathbf{h}(k)$ into
$10$ clusters. Without the RBM, the same clustering method extracts $12$
clusters from the original data $\mathbf{x}(k)$.

\item Membership functions training. In order to improve modeling accuracy,
the membership functions of the fuzzy model are updated with the input and
output data. The centers of the membership function are the cluster centers
which are obtained in Step $2$. The parameters $W$ are computed using the ELM
approach (we calculate the pseudoinverse using a vector which contains the
parameters $\Phi$).

\item Probability training. Once the minimization problem is set, The
probability parameters $p_{i,j}$ are estimated by the standard linear
programming toolbox, "$\mathrm{fmincon"}$ and "\textrm{sqp"}. The initial
value for the matrix $P$ is the identity matrix $\mathbf{I}_{K},$ \textit{i.
e.}, we start from a standard fuzzy rule and the probability parameters are
introduced to minimize the possibility of the modeling error procedure.
\end{enumerate}

\begin{figure}
[ptb]
\begin{center}
\includegraphics[
natheight=4.629900in,
natwidth=8.141700in,
height=3.0459in,
width=5.3333in
]%
{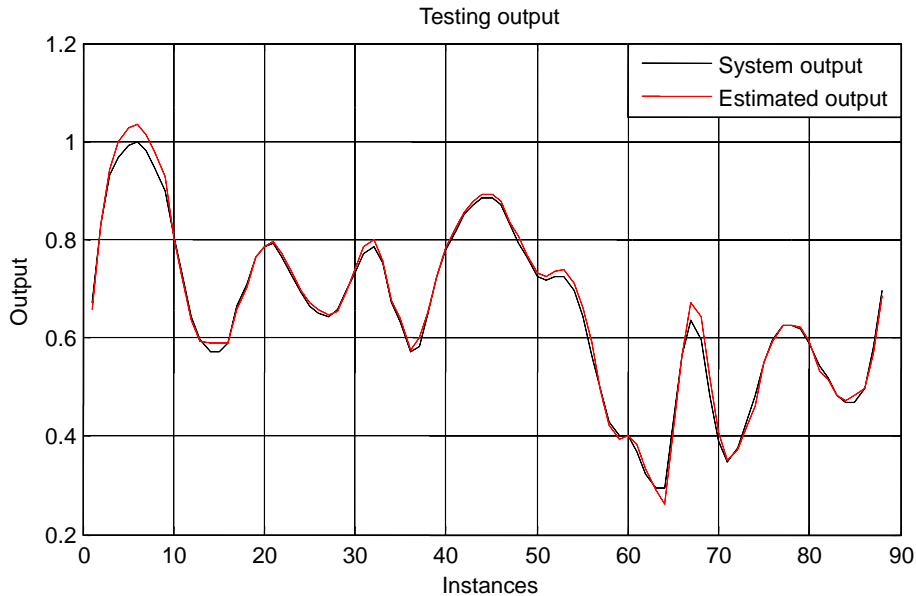}%
\caption{Testing results of the gas furnace modeling.}%
\label{FigOutGas}%
\end{center}
\end{figure}

In order to test the generalization capabilities of our model, we use the
remaining $96$ data for testing after the training phase is finished. The
final testing results are shown in Figure \ref{FigOutGas}.%

%TCIMACRO{\FRAME{ftbpFU}{5.3333in}{3.0459in}{0pt}{\Qcb{Testing results of the
%gas furnace modeling.}}{\Qlb{FigOutGas}}{Figure}%
%maintain-aspect-ratio TRUE;  display "USEDEF";  valid_file "T";
%width 5.3333in;  height 3.0459in;  depth 0pt;  original-width 8.1417in;
%original-height 4.6299in;  cropleft "0";  croptop "1";  cropright "1";
%cropbottom "0";  tempfilename 'TestGas.eps';tempfile-properties "XNPR";}} }%
%BeginExpansion
%EndExpansion

We compared our method with the following three fuzzy modeling algorithms:

\begin{enumerate}
\item Adaptive fuzzy modeling approach (ANFIS) \cite{Juangb}\cite{Wang}. It
may be the most popular fuzzy modeling method. In this experiment, we also use
$8$ fuzzy rules. The Gaussian membership functions are selected randomly at first.

\item Fuzzy modeling via online clustering \cite{Juang2}\cite{Tzafestas}%
\cite{Angelov}. Here we do not consider the temporal interval problem
\cite{Yu1} and use all data to train each group. All thresholds for the output
and the input are $1.5$. Finally ,we obtain five fuzzy rules.

\item Fuzzy logic with data clustering \cite{Mitra}\cite{Chiu}. It is another
popular fuzzy modeling method. In this comparison, only the input is
partitioned. With the threshold $1.0$, we have $15$ groups in the input space.
So $15$ fuzzy rules are constructed.
\end{enumerate}

The root mean square (RMS) testing error for each method is $RMS_{1}=0.019$
(our fuzzy modeling with RBMs), $RMS_{2}=0.031$ (fuzzy modeling with
clustering) and $RMS_{3}=0.09$ (ANFIS).

\begin{figure}
[ptb]
\begin{center}
\includegraphics[
natheight=5.635900in,
natwidth=7.511700in,
height=2.9153in,
width=5.2278in
]%
{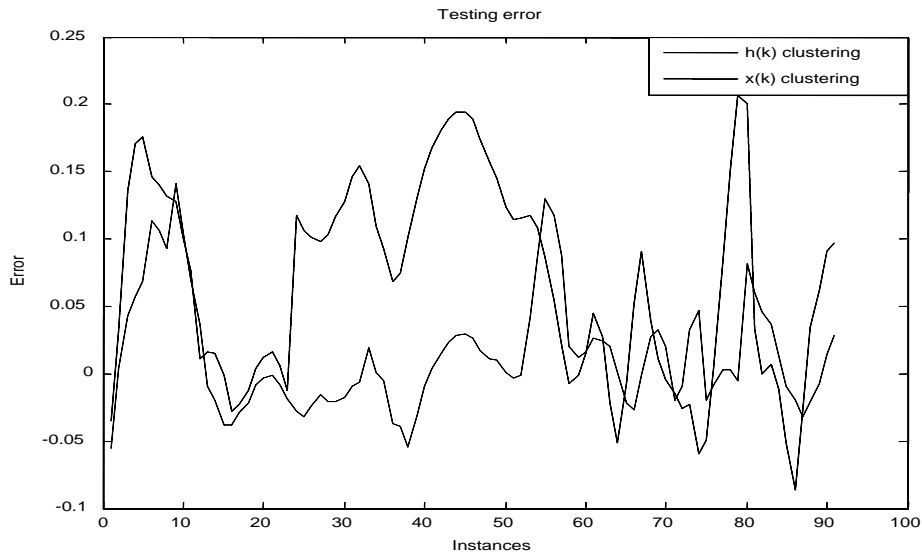}%
\caption{Testing errors with RBMs and without RBMs}%
\label{FigTestGas}%
\end{center}
\end{figure}

In order to show the effectiveness of the hidden feature extraction with RBMs,
we compare the testing error of $\mathbf{h}(k)$ clustering after RBMs and
$\mathbf{x}(k)$ clustering without RBMs. Figure \ref{FigTestGas} gives these
testing errors.

It is observed that the clustering procedure using the features from the RBM
gives better representation for the input data. Once the fuzzy rules are
trained, the hidden features can be observed by the RBM, and the probabilistic
fuzzy model improve the modeling accuracy.%

%TCIMACRO{\FRAME{ftbpFU}{5.2278in}{2.9153in}{0pt}{\Qcb{Testing errors with RBMs
%and without RBMs}}{\Qlb{FigTestGas}}{Figure}%
%{\special{ language "Scientific Word";  type "GRAPHIC";  display "USEDEF";
%valid_file "T";  width 5.2278in;  height 2.9153in;  depth 0pt;
%original-width 7.5117in;  original-height 5.6359in;  cropleft "0";
%tempfilename 'ErrorGas1.eps';tempfile-properties "XNPR";}} }%
%BeginExpansion
%EndExpansion

\begin{figure}
[ptb]
\begin{center}
\includegraphics[
natheight=5.635900in,
natwidth=7.511700in,
height=2.8245in,
width=5.2486in
]%
{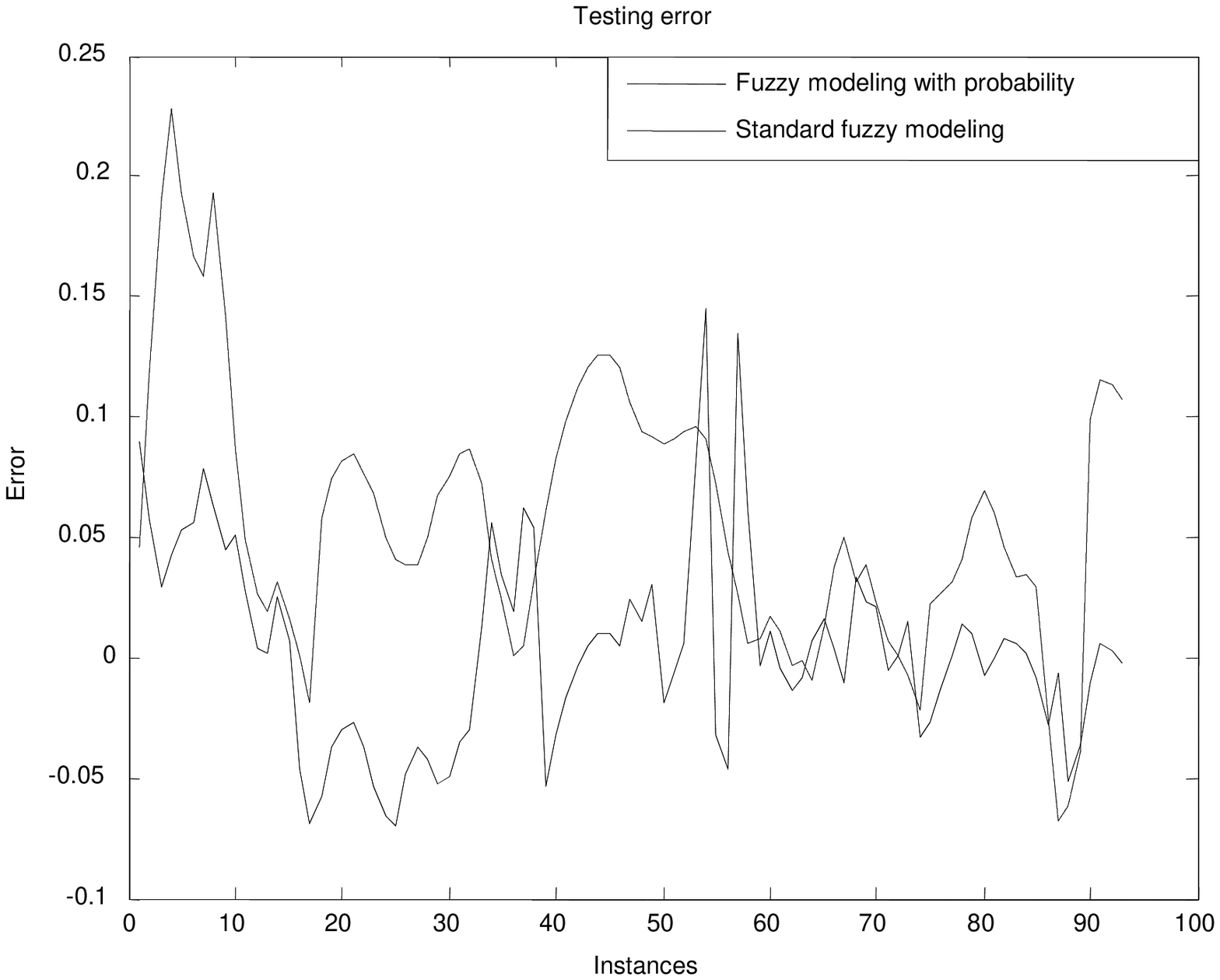}%
\caption{GAS testing error using probabilistic parameters}%
\label{FigTestGas2}%
\end{center}
\end{figure}

Now we discuss how the probability parameters work in the consequences of the
fuzzy rules (\ref{frp}). Figure \ref{FigTestGas2} shows the training errors
with standard fuzzy rules and probabilistic fuzzy rules. We see that the
probabilistic parameters give more freedom and robustness to adjust the model
with the data, the testing errors decrease in the most of time.%

%TCIMACRO{\FRAME{ftbpFU}{5.2486in}{2.8245in}{0pt}{\Qcb{GAS testing error using
%probabilistic parameters}}{\Qlb{FigTestGas2}}{Figure}%
%{\special{ language "Scientific Word";  type "GRAPHIC";  display "USEDEF";
%valid_file "T";  width 5.2486in;  height 2.8245in;  depth 0pt;
%original-width 7.5117in;  original-height 5.6359in;  cropleft "0";
%tempfilename 'ErrorGas2.eps';tempfile-properties "XNPR";}} }%
%BeginExpansion
%EndExpansion

The mean square errors (MSE) of using RBMs for the clustering and probability
parameters for the fuzzy rules are given in Table 1. We see how the use of
each stage clearly helps with the decreasing of the testing error.

\begin{center}
Table 1. MSE of the gas furnace modeling ($\times10^{-3})$%

\begin{tabular}
[c]{c|c|c|c|c}\cline{2-2}\cline{4-4}
& Training &  & Testing & \\\cline{2-2}\cline{2-5}\cline{4-4}
& No RBM & RBM & No RBM & \multicolumn{1}{|c|}{RBM}\\\hline
\multicolumn{1}{|c|}{{\small Standard fuzzy rule}} & $5.10$ & $3.35$ & $26.2$
& \multicolumn{1}{|c|}{$23.7$}\\\hline
\multicolumn{1}{|c|}{{\small Probabilistic fuzzy rule}} & $3.25$ & $3.11$ &
$22.5$ & \multicolumn{1}{|c|}{$19.3$}\\\hline
\end{tabular}

\end{center}

\subsection{Wiener-Hammerstein benchmark problem}

Wiener-Hammerstein (W-H) system is series connection of three parts: a linear
system, a static nonlinearity and other independent linear system. The data of
the Wiener-Hammerstein benchmark is generated from an electrical circuit which
consists in the above cascade blocks \cite{Schoukens}. There is not direct
measurement to the static nonlinearity, because it is located between two
unknown linear dynamic systems.

The benchmark data set consists $188,000$ input/output pairs. The data set is
divided in two parts: $100,000$ sample pairs are for training and $88,000$
samples are for testing.

Let $u(k)$ be the input and $y\left(  k\right)  $ be the output. We define the
recursive input vector to the model as $\mathbf{x}(k)=\left[  y(k-1)\cdots
y(k-n_{y})\text{ }u(k)\cdots u(k-n_{u})\right]  ^{T}.$ So the
Wiener-Hammerstein benchmark is%
\begin{equation}
y\left(  k\right)  =f\left[  y(k-1)\cdots y(k-n_{y})\text{ }u(k)\cdots
u(k-n_{u})\right]  \label{wh}%
\end{equation}
Similar as the previous example, $u(k)$ and $y\left(  k\right)  $ are also
normalized. The delays $n_{y}$ and $n_{u}$ are drawn again from a uniform
interval $[1,10].$ The fuzzy modeling process also has the following four steps:

\begin{enumerate}
\item Features extraction. We also train the RBM with contrastive divergence
with $1$-step Gibbs sampling and $10$ training epochs. The learning rate is
$\eta=0.1$. Due to the quantity of data, we utilize the lesser learning rate.
The number of hidden units is also chosen as $n_{y}+n_{u}+1$.

\item Clustering. We set $\alpha=0.95$, $\psi=100,$ $\lambda=5$ and $C=0.001$.
$\alpha$ and $\psi$ determine the probability given by the Chinese restaurant
process, $\alpha$ is chosen close to 1 to ensure that a big number of clusters
are created, $\psi$ also increases to accomplish the same objective. The
hidden feature $\mathbf{h}(k)$ is divided into $13$ clusters, while the
original data $\mathbf{x}(k)$ is partitioned into $11$ clusters.

\item Membership functions training. The parameters $W$ are again computed
using the pseudoinverse approach.

\item Probability training. We used Matlab functions: $\mathrm{fmincon}$ and
\textrm{sqp}, to compute the parameters $p_{i,j}$. $P$ is initialized as
$\mathbf{I}_{K}.$
\end{enumerate}

%

%TCIMACRO{\FRAME{ftbpFU}{5.303in}{2.9084in}{0pt}{\Qcb{Data-driven fuzzy
%modeling method for the W-H data}}{\Qlb{FigFolWL}}{Figure}%
%{\special{ language "Scientific Word";  type "GRAPHIC";  display "USEDEF";
%valid_file "T";  width 5.303in;  height 2.9084in;  depth 0pt;
%original-width 7.9546in;  original-height 4.0124in;  cropleft "0";
%tempfilename 'testWH.eps';tempfile-properties "XNPR";}} }%
%BeginExpansion
\begin{figure}
[ptb]
\begin{center}
\includegraphics[
natheight=4.012400in,
natwidth=7.954600in,
height=2.9084in,
width=5.303in
]%
{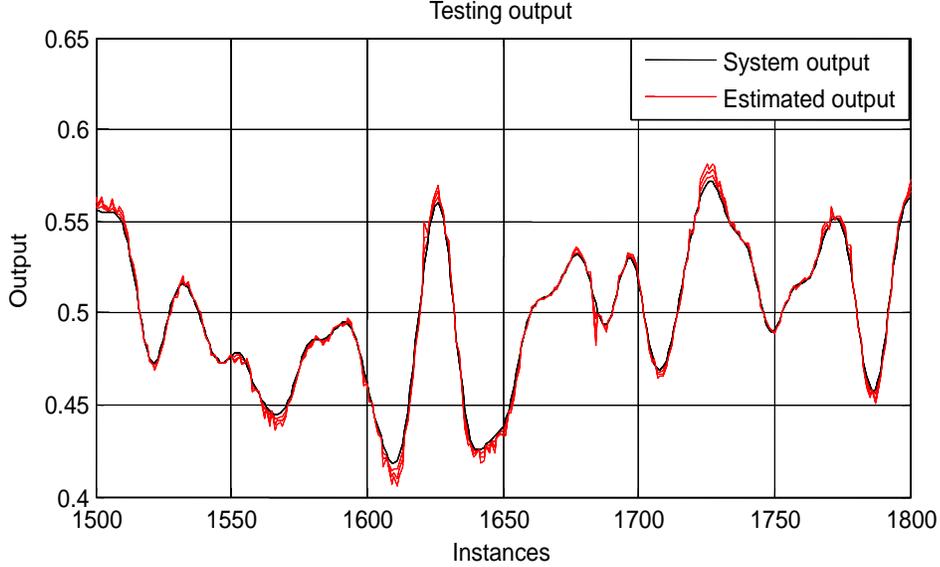}%
\caption{Data-driven fuzzy modeling method for the W-H data}%
\label{FigFolWL}%
\end{center}
\end{figure}
%EndExpansion

\begin{figure}
[ptb]
\begin{center}
\includegraphics[
natheight=5.635900in,
natwidth=7.511700in,
height=2.936in,
width=5.5469in
]%
{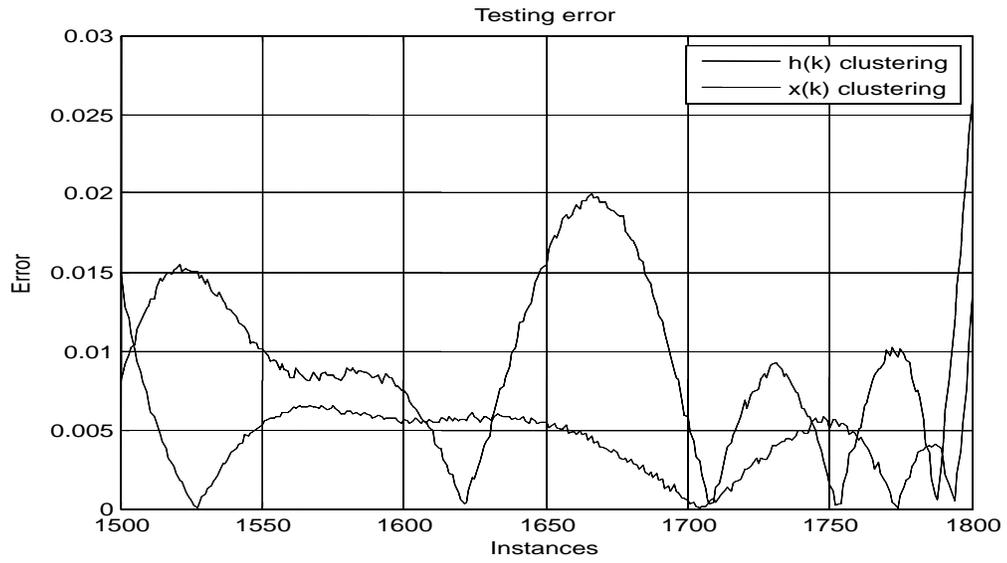}%
\caption{Testing errors using RBM and without RBM}%
\label{FigTestWH}%
\end{center}
\end{figure}

Our data-driven fuzzy modeling method for the W-H data is shown in Figure
\ref{FigOutGas}. To see how the RBM to help to decrease the modeling error,
Figure \ref{FigTestWH} shows the testing errors for $\mathbf{x}(k)$ and
$\mathbf{h}(k)$ clustering. We can see that clustering directly over
$\mathbf{x}(k)$ gives a good testing performance but its MSE is greater.%

%TCIMACRO{\FRAME{ftbpFU}{5.5469in}{2.936in}{0pt}{\Qcb{Testing errors using RBM
%and without RBM}}{\Qlb{FigTestWH}}{Figure}%
%{\special{ language "Scientific Word";  type "GRAPHIC";  display "USEDEF";
%valid_file "T";  width 5.5469in;  height 2.936in;  depth 0pt;
%original-width 7.5117in;  original-height 5.6359in;  cropleft "0";
%tempfilename 'FigErrorWH.eps';tempfile-properties "XNPR";}} }%
%BeginExpansion
%EndExpansion

\begin{figure}
[ptb]
\begin{center}
\includegraphics[
natheight=5.635900in,
natwidth=7.511700in,
height=2.866in,
width=5.2693in
]%
{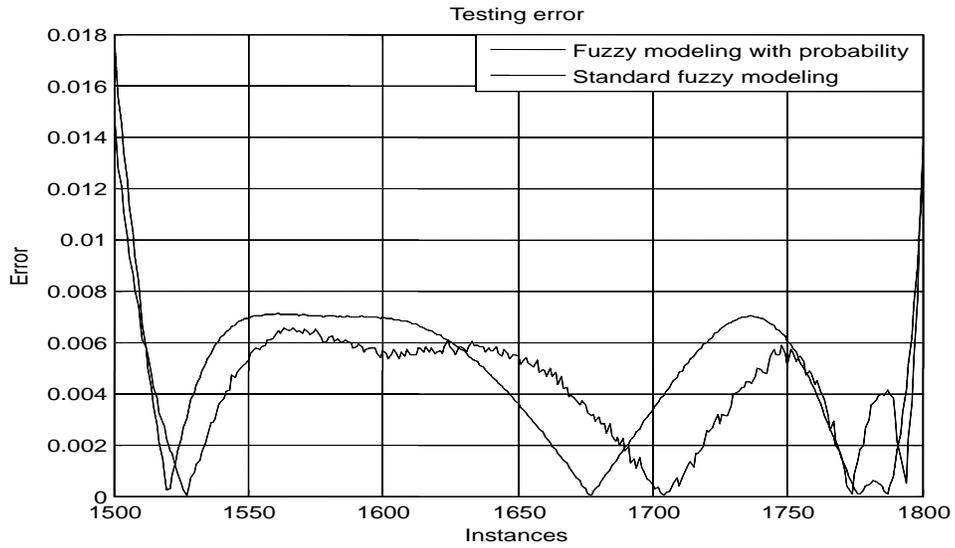}%
\caption{Testing errors using probabilistic parameters and standard fuzzy
rules}%
\label{FigTestWH2}%
\end{center}
\end{figure}

Figure \ref{FigTestWH2} shows the effect of the fuzzy probability parameters
$p_{i,j}$. We see that as the number of clusters $K$ increased the
computational time of the model decreased, this is due to the linear
programming method for calculation of $P$. The MSE decreases when we use
probabilistic fuzzy rules.%

%TCIMACRO{\FRAME{ftbpFU}{5.2693in}{2.866in}{0pt}{\Qcb{Testing errors using
%probabilistic parameters and standard fuzzy rules}}{\Qlb{FigTestWH2}}%
%{Figure}{\special{ language "Scientific Word";  type "GRAPHIC";
%display "USEDEF";  valid_file "T";  width 5.2693in;  height 2.866in;
%depth 0pt;  original-width 7.5117in;  original-height 5.6359in;
%cropleft "0";  croptop "1";  cropright "1";  cropbottom "0";
%tempfilename 'FigTest2WH.eps';tempfile-properties "XNPR";}} }%
%BeginExpansion
%EndExpansion

By combining the restricted Boltzmann machines and the probability theory, our
data-driven fuzzy modeling method has outstanding property, see Table 2.

\begin{center}
Table 2. MSE of W-H modeling ($\times10^{-3})$%

\begin{tabular}
[c]{c|c|c|c|c}\cline{2-2}\cline{4-4}
& Training &  & Testing & \\\cline{2-2}\cline{2-5}\cline{4-4}
& No RBM & RBM & No RBM & \multicolumn{1}{|c|}{RBM}\\\hline
\multicolumn{1}{|c|}{{\small Standard fuzzy rule}} & $18.9$ & $17.7$ & $26.4$
& \multicolumn{1}{|c|}{$22.8$}\\\hline
\multicolumn{1}{|c|}{{\small Probabilistic fuzzy rule}} & $16.2$ & $14.1$ &
$23.6$ & \multicolumn{1}{|c|}{$19.3$}\\\hline
\end{tabular}

\end{center}

We find that the modeling accuracy of the W-H benchmark does not improve so
much as the gas furnace by the probabilistic tuning. While the RBM gives
better results when more data are available.

\section{Conclusions}

In this paper we propose an efficient data-driven modeling approach for
nonlinear system modeling using fuzzy rules. Several techniques are applied to
the fuzzy modeling. We propose a modified restricted Boltzmann machine to
extract hidden features. A probabilistic clustering method\ is designed to
partition the input and output data into several clusters. After the structure
identification, we apply ELM to train the consequent part of the fuzzy rules,
while the parameters of the premise part come from the probabilistic
clustering directly. Finally, the probability parameters are introduced into
the fuzzy rules to enhance the expression capabilities of the model.

Our method can be extended to online modelling of nonlinear systems by using
online clustering with RBM\ and the adaptive fuzzy modeling techniques. Since
we use deep learning pre-training, the probabilistic clustering, and linear
programming for the probabilistic parameters, the computational time is longer
but the accuracy is improved significantly.

\end{document}